\documentclass[twocolumn]{aastex62}
\newcommand{\logg}{$\log\,g$}
\newcommand{\teff}{$T_{\rm eff}$}

\received{XXX, 2018}
\revised{YYY, 2018}
\accepted{ZZZ, 2018}
\submitjournal{ApJL}

\shorttitle{Weak Magnetic Fields of AK\,Sco and HD\,95881}
\shortauthors{J\"arvinen et al.}

\begin{document}

\title{Weak Magnetic Fields in Two Herbig~Ae Systems:
the SB2 AK\,Sco and the Presumed Binary HD\,95881}


\correspondingauthor{S.~P.~J\"arvinen}
\email{sjarvinen@aip.de}

\author[0000-0003-3572-9611]{S.~P.~J\"arvinen}
\affiliation{Leibniz-Institut f\"ur Astrophysik Potsdam (AIP),
  An der Sternwarte~16, 14482~Potsdam, Germany}

\author{T.~A.~Carroll}
\affiliation{Leibniz-Institut f\"ur Astrophysik Potsdam (AIP),
  An der Sternwarte~16, 14482~Potsdam, Germany}

\author[0000-0003-0153-359X]{S.~Hubrig}
\affiliation{Leibniz-Institut f\"ur Astrophysik Potsdam (AIP),
  An der Sternwarte~16, 14482~Potsdam, Germany}

\author{I.~Ilyin}
\affiliation{Leibniz-Institut f\"ur Astrophysik Potsdam (AIP),
  An der Sternwarte~16, 14482~Potsdam, Germany}

\author[0000-0002-5379-1286]{M.~Sch\"oller}
\affiliation{European Southern Observatory,
  Karl-Schwarzschild-Str.~2, 85748 Garching, Germany}

\author{F.~Castelli}
\affiliation{Istituto Nazionale di Astrofisica,
  Osservatorio Astronomico di Trieste, via Tiepolo 11, 34143, Trieste, Italy}

\author[0000-0002-4308-0763]{C.~A.~Hummel}
\affiliation{European Southern Observatory,
  Karl-Schwarzschild-Str.~2, 85748 Garching, Germany}

\author{M.~G.~Petr-Gotzens}
\affiliation{European Southern Observatory,
  Karl-Schwarzschild-Str.~2, 85748 Garching, Germany}

\author[0000-0003-0529-1161]{H.~Korhonen}
\affiliation{Dark Cosmology Centre, Niels Bohr Institute,
  University of Copenhagen, Juliane Maries Vej~30, 2100 Copenhagen, Denmark}

\author[0000-0001-9754-2233]{G.~Weigelt}
\affiliation{Max-Planck Institut f\"ur Radioastronomie,
  Auf dem H\"ugel~69, 53121 Bonn, Germany}

\author{M.~A.~Pogodin}
\affiliation{Central Astronomical Observatory at Pulkovo of Russian Academy
  of Sciences, 196140 Saint Petersburg, Russia}

\author[0000-0003-4842-8834]{N.~A.~Drake}
\affiliation{Observat\'orio Nacional/MCTIC,
  Rua Gen. Jos\'e Cristino, 77, 20921-400 Rio de Janeiro, Brazil}
\affiliation{Laboratory of Observational Astrophysics, Saint Petersburg State
  University, Universitetskij pr.~28, 198504 Saint Petersburg, Russia}


\begin{abstract}

  We report the detection of weak mean longitudinal magnetic fields in the
  Herbig\,Ae double-lined spectroscopic binary AK\,Sco and in the presumed
  spectroscopic Herbig\,Ae binary HD\,95881 using observations with
  the High Accuracy Radial velocity Planet Searcher polarimeter (HARPSpol)
  attached to the European Southern Observatory’s (ESO’s) 3.6\,m
  telescope. Employing a multi-line singular value decomposition (SVD)
  method, we detect a mean longitudinal magnetic field
  $\left<B_{\mathrm z}\right>=-83\pm31$~G in the secondary component
  of AK\,Sco on one occasion. For HD\,95881, we measure
  $\left<B_{\mathrm z}\right>=-93\pm25$~G and
  $\left<B_{\mathrm z}\right>=105\pm29$~G at two different observing epochs. For
  all the detections the false alarm probability is smaller than $10^{-5}$.
  For AK\,Sco system, we discover that accretion diagnostic
  Na\,{\footnotesize I} doublet lines and photospheric lines show intensity
  variations over the observing nights. The double-lined spectral appearance
  of HD\,95881 is presented here for the first time.
  
\end{abstract}


\keywords{
  binaries: spectroscopic ---
  stars: abundances ---
  stars: individual (AK\,Sco, HD\,95881) ---
  stars: magnetic field ---
  stars: pre-main sequence}


\section{Introduction} \label{sec:intro}

Studies of the presence of magnetic fields in Herbig\,Ae/Be stars are
extremely important because they enable us to improve our insight into how
the magnetic fields of these stars are generated and how they interact with
their environment, including their impact on the planet formation processes
and the planet--disk interaction. Fourteen years ago,
\citet{Muzerolle}
confirmed magnetospheric accretion Balmer and sodium line profiles in the
Herbig\,Ae star UX\,Ori. Since then, a number of magnetic studies have
been attempted, indicating that about 20 Herbig\,Ae/Be stars likely have
globally organized magnetic fields 
\citep[][and references therein]{Hubrig2015}.
While Herbig\,Ae stars show evidence for magnetospheric accretion, it was
suggested that Herbig\,Be stars might accrete directly from a boundary layer
instead of the magnetosphere
\citep[e.g.,][]{HamannPersson, GuentherSchmitt}.
Interestingly, the compilation of existing magnetic field detections in
Herbig\,Ae/Be stars by
\citet{Hubrig2015}
showed that only very few Herbig\,Be stars, all of them of spectral type
B9, possess mean longitudinal magnetic fields.

Because only about 20 Herbig\,Ae/Be stars have been reported to possess
magnetic fields, several arguments have recently been presented that favor a
scenario in which the low detection rate of magnetic fields of Herbig\,Ae
stars can be explained by the weakness of these fields and rather large
measurement uncertainties
\citep{Hubrig2015}.
The obtained density distribution of the rms longitudinal
magnetic field values revealed that only a few stars have fields stronger
than 200\,G, and half of the sample possess magnetic fields of about 100\,G
or less. These results call into question our current understanding of the
magnetospheric accretion process in intermediate-mass pre-main-sequence stars
as they indicate that the magnetic fields of Herbig\,Ae/Be stars are by far
weaker than those measured in their lower-mass classical T\,Tauri star
counterparts, usually possessing kG magnetic fields. We note that the task of
detecting weak magnetic fields of Herbig stars is very challenging due to the
complex interaction between the stellar magnetic field, the accretion disk,
and the stellar wind. Therefore, to analyze the presence of weak stellar
magnetic fields, we usually employ a dedicated software package, the so-called
multi-line singular value decomposition (SVD) method for Stokes profile
reconstruction developed by
\citet{SVD}.

In this Letter, we present our search for a magnetic field in the close
double-lined spectroscopic binary with a Herbig\,Ae component AK\,Sco and
in the presumed Herbig\,Ae binary system HD\,96881 using high-resolution
spectropolarimetric observations obtained with the High Accuracy Radial
velocity Planet Searcher polarimeter
\citep[HARPSpol;][]{snik2008}
attached to the European Southern Observatory’s (ESO’s)
3.6\,m telescope (La Silla, Chile). Herbig Ae/Be binaries
are of special interest as no close double-lined spectroscopic binary with a
magnetic Herbig Ae component was detected among the 20 Herbig stars known
to possess magnetic fields. AK\,Sco is a ninth-magnitude close SB2 system
($P_{\rm orb}=13.6$\,days) with approximately equal components surrounded by a
circumbinary disk
\citep{Alencar2003}
and classified as a class II Herbig\,Ae/Be system by
\citet{Menu2015}.
A study of this actively accreting system is of particular importance
because of its prominent ultraviolet excess
and the high eccentricity ($e=0.47$) of its orbit -- the components get as
close as 11\,$R_{*}$ at periastron passage. It was suggested by
\citet{Gomez2013}
that the strong UV emission is caused by reprocessing of the high-energy
magnetospheric radiation by the circumstellar material and that the eccentric
orbit acts like a gravitational piston.

Whereas  AK\,Sco was intensively studied in the last few years, there is little
information available on the eighth-magnitude target HD\,95881 in the literature.
Based on spectro-astrometric observations,
\citet{Baines2006}
considered HD\,95881 as a possible sub-arcsecond binary. The authors report a
displacement of the FWHM over the H$\alpha$ profile by
0.02\,arcsec, consistent with a binary detection.
\citet{Verhoeff2010}
mapped the spatial distribution of the gas and dust around HD\,95881 and
concluded that it is in the transition phase from a gas-rich flaring dust
disk to a gas-poor self-shadowed disk. The double-lined spectral appearance of
HD\,95881 is presented here for the first time.


\section{Observations and Magnetic Field Analysis}
\label{sec:obs}

\begin{deluxetable}{lcclc}[t!]
  \tablecaption{Logbook of AK\,Sco and HD\,95881 HARPSpol observations \label{tab:log}}
\tablewidth{0pt}
\tablehead{
\colhead{AK\,Sco} & \colhead{} & \colhead{} & \colhead{HD\,95881} & \colhead{}\\
  \colhead{HJD} & \colhead{S/N} & \colhead{Phase} & \colhead{HJD} & \colhead{S/N} \\
\colhead{2,400,000+} & \colhead{} & \colhead{} & \colhead{2,400,000+} & \colhead{}
}
\startdata
57554.798230 & 144 & 0.946 & 57554.588158 & 193 \\
57555.782879 & 195 & 0.018 & 57908.557375 & 184 \\
57908.706901 & 224 & 0.950 & 57909.538571 & 260 \\
57909.716316 & 282 & 0.025 & 57910.551552 & 219 \\
57910.605621 & 205 & 0.090 & 57911.546578 & 203 \\
57911.684945 & 207 & 0.169 & {          } & { } \\
\enddata
\tablecomments{The columns give the heliocentric Julian date (HJD) for the
  middle of the exposures and the signal-to-noise ratio (S/N) of the spectra.
  The orbital phase of AK\,Sco is based on the period
  $P=13.609453\pm 0.000026$~days and periastron passage time
  $T=2,446,654.3634\pm 0.0086$
  \citep{Alencar2003}.}
\end{deluxetable}

HARPSpol spectropolarimetric observations of AK\,Sco and HD\,95881 were
obtained in the nights of 2016 June 15 and 16, and on four consecutive nights
between 2017 June 4 and 7. Each observation consisted of subexposures with
exposure times varying between 30 and 48 minutes for AK\,Sco and between 38 and
45 minutes for HD\,95881. The quarter-wave retarder plate was rotated by
$90\degr$ after each subexposure. All spectra have a resolving power of about
110,000 and cover the spectral range 3780--6910\,\AA{}, with a small gap
between 5259 and 5337\,\AA{}. The reduction and calibration of these
spectra was performed using the HARPS data reduction software available on
La~Silla. The normalization of the spectra to the continuum level is described
in detail by
\citet{Hubrig2013}.
The summary of our HARPSpol observations is given in Table~\ref{tab:log}. For
AK\,Sco, the orbital phases were calculated using the orbital period
$P=13.609453\pm 0.000026$~days and the epoch of periastron
$T=2\,446\,654.3634\pm 0.0086$
\citep{Alencar2003}.
At four orbital phases, from 0.950 to 0.090, the spectral lines of both
components in the spectra of AK\,Sco are well separated. Only at the phase
0.169, corresponding to conjunction time, are the spectral lines of
both components fully overlapped. The noisy HARPS observation obtained
in 2015  June 15 at the orbital phase $\phi$=0.946 shows strong
contamination, where the secondary component is hardly detectable
probably due to the presence of clumpy dust clouds in the circumbinary
disk and/or inside the orbit of the system. We discuss the appearance
of this spectrum in more detail in Section~\ref{sec:mfield}.

\begin{figure}[ht!]
\includegraphics[width=\columnwidth]{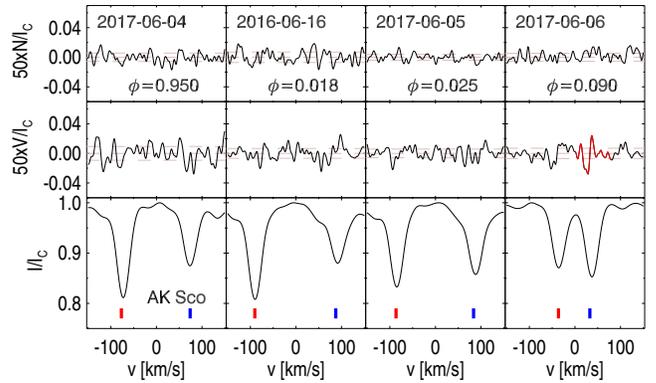}
\caption{SVD Stokes~$I$ (bottom), $V$ (middle), and diagnostic null (N)
  profiles (top) obtained for AK\,Sco. The Stokes~$V$ and N profiles have
  been amplified by factor 50. The detected Zeeman feature is highlighted in
  red. The profiles are sorted according to the orbital phase. Under the
  Stokes~$I$ profiles, the components are marked with red (primary component)
  and blue (secondary component) ticks. The horizontal dashed lines indicate
  the average values and the $\pm 1\sigma$-ranges.
  \label{fig:akmag}}
\end{figure}

\begin{figure}[ht!]
\includegraphics[width=\columnwidth]{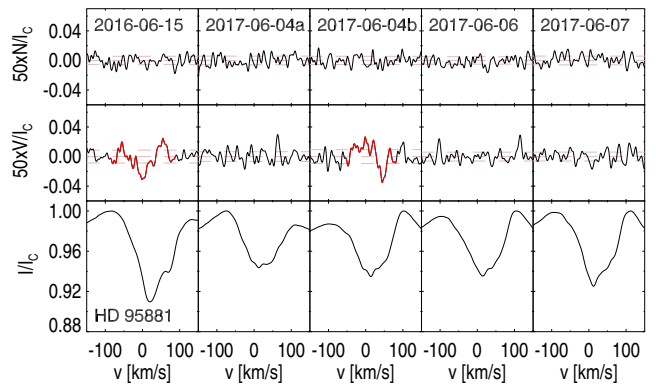}
\caption{Same as in Figure~\ref{fig:akmag} but for HD\,95881. The profiles are plotted
  corresponding to the increasing date. The detected Zeeman features are
  highlighted with red.
  \label{fig:hd95mag}}
\end{figure}

The magnetic field analysis of both systems is done based on the SVD
method developed by
\citet{SVD}.
The SVD method is similar to the Principle Component Analysis approach
\citep[see also][]{CarrollKopf, MGetal2008, Semel2009},
where the similarity of the individual Stokes~$V$ profiles allows one to
describe the most coherent and systematic features present in all spectral
line profiles as a projection onto a small number of eigenprofiles
\citep{SVD}.
This method is most useful for analyzing the presence of weak
stellar magnetic fields and has been proven to achieve accuracies of just a few Gauss
\citep[e.g.,][]{Hubrig2015}.
The SVD Stokes $I$ and $V$ spectra for AK\,Sco and HD\,95881 are calculated
using the line masks based on their atmospheric parameters
\citep{Alencar2003, Fairlamb2015}
and were constructed  using the Vienna Atomic Line Database
\citep[VALD;][]{Kupka2011, VALD3}.
For the study of AK\,Sco, we have selected 50 spectral lines, and 
for HD\,95881 32. These lines are blend-free and belong to iron-peak
elements. The SVD Stokes $I$ and $V$ spectra for both systems are then
used for the measurement of the longitudinal magnetic field through
the first-order moment of the Stokes $V$ profile
\citep[e.g.,][]{Mathys1989},
assuming the mean Land\'e factor $\bar{g}_{\rm eff} = 1.192$ for AK\,Sco and
$\bar{g}_{\rm eff} = 1.186$ for HD\,95881. The Stokes~$I$, $V$, and diagnostic
$N$ SVD profiles for the system AK\,Sco obtained using HARPSpol observations
distributed over the orbital phases 0.950--0.090 are presented in
Figure~\ref{fig:akmag}, and those for HD\,95881 are presented in 
Figure~\ref{fig:hd95mag}. The diagnostic $N$ profiles are usually used
to identify spurious polarization signatures. They are calculated by
combining the subexposures in such a way that the polarization cancels out.

For the system AK\,Sco we observe in the Stokes $V$ spectra a number of small
signatures corresponding to the position of the components, but a definite
magnetic field detection, $\left<B_{\mathrm z}\right>=-83\pm31$\,G, with a
false alarm probability (FAP) smaller than $10^{-5}$ was achieved only for the
secondary component component at the orbital phase $\phi = 0.090$. We classify
the magnetic field measurements making use of the FAP
\citep{Donati-fap},
considering a profile with FAP\,$<$\,$10^{-5}$ as a definite detection,
$10^{-5}$\,$<$\,FAP\,$<10^{-3}$ as a marginal detection, and
FAP\,$>$\,$10^{-3}$ as a non-detection. According to
\citet{Alencar2003}
both components are expected to be tidally synchronized, so we observe at this
phase the region of the stellar surface facing permanently the primary
component, meaning that the magnetic field geometry in this component is likely
closely related to the position of the primary component. We note that a
similar magnetic field behavior, where the field orientation is linked to the
companion, was previously detected in two other (the only known) close
main-sequence binaries with Ap components, HD\,98088 and HD\,161701
\citep{Babcock, Hubrig2014}.
We do not detect a magnetic field in the primary, but as our observations
cover only a fraction of the orbital cycle, we cannot exclude the possibility
that the primary also possesses a magnetic field. In such a case, both stars
could undergo common magnetospheric accretion. This idea is supported by
\emph{Hubble Space Telescope} observations indicating the presence of unresolved
macroscopic motions in the dense and warm circumstellar material surrounding
this system
\citep{Gomezet}.

Our HARPSpol spectra of HD\,95881 clearly show the
presence of two components that remain blended with each other on all five
observing nights. Definite detections with FAP $ < 10^{-5}$ are achieved from
the SVD profiles on two observing epochs, with the measured magnetic field
strengths $\left<B_{\mathrm z}\right>=-93\pm25$~G and
$\left<B_{\mathrm z}\right>=105\pm29$~G, respectively. As mentioned in
Section~\ref{sec:intro},
\citet{Baines2006}
used long-slit spectra for a spectro-astrometric analysis of HD\,95881 and
suggested it as a possible binary. Because the binary nature of this target is
not convincingly ascertained, we cannot exclude that the variability of the
shape of the SVD Stokes~$I$ profile is caused by a contribution of the 
circumstellar material around the star. Long-term spectroscopic monitoring is 
certainly needed to be able to separate the photospheric and non-photospheric
contributions in the line profiles of HD\,95881 and to come to a conclusion on the
binarity status.


\section{Spectral Variability}
\label{sec:mfield}

The preliminary abundance analysis of AK\,Sco performed by directly fitting
synthetic spectra to the observations at phase $\phi$=0.090, where the weak
magnetic field was discovered for the secondary, revealed Sr and Ba
overabundances of 0.5\,dex in the secondary and a Sr underabundance of
0.4\,dex in the primary. Strong Li absorption lines indicate an overabundance
of 2.2\,dex in the primary and 2.7\,dex in the secondary. We used for both
components the ATLAS9 model with \teff=6500\,K
\citep{Alencar2003},
\logg=4.0, and microturbulent velocity $\xi$=2\,km\,s$^{-1}$. Synthetic
spectra of both stars were computed with the SYNTHE code
\citep{synthe}.

\begin{figure}[ht!]
\includegraphics[width=\columnwidth]{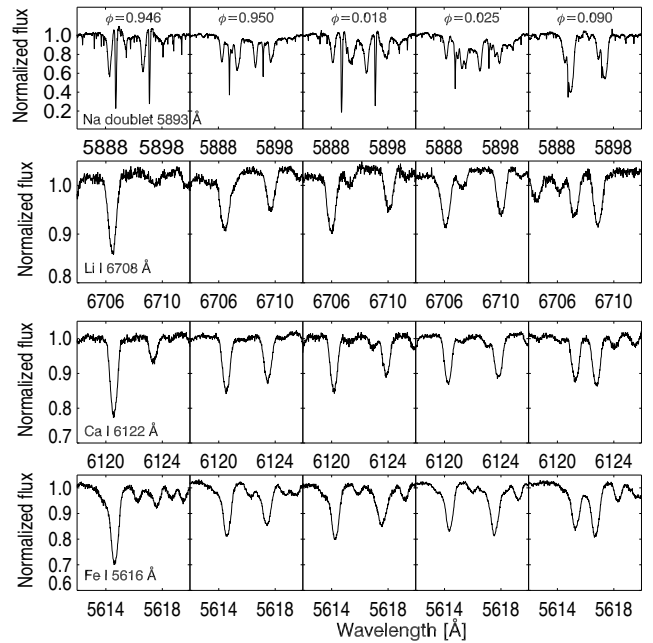}
\caption{From bottom to top: variability of the spectral lines
  Fe\,{\footnotesize I}~5616, Ca\,{\footnotesize I}~6122,
  Li\,{\footnotesize I}~6708, and the accretion diagnostic
  Na\,{\footnotesize I} doublet in AK\,Sco.
  \label{fig:akelem}}
\end{figure}

\begin{figure}[ht!]
\includegraphics[width=\columnwidth]{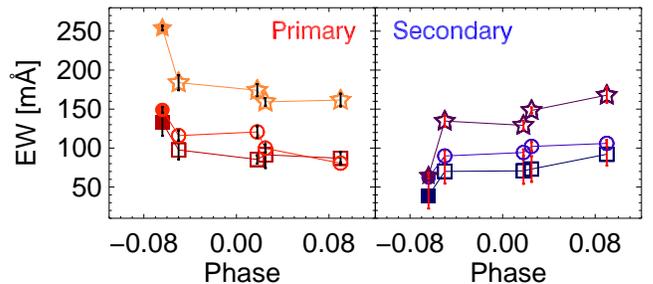}
\caption{Orbital phase evolution of equivalent width of lithium (squares),
  calcium (circles), and iron (stars) lines for the primary (left side) and
  for the secondary (right side) components of AK\,Sco. The filled symbols
  denote the first data set obtained in 2016. The measurement error bars of
  the order of 3--17\,m\AA{} are indicated inside the symbols.
  \label{fig:akew}}
\end{figure}

We discovered that all accretion diagnostic lines and photospheric lines show
intensity variations over the observing nights. In view of the complexity of
this system, it is impossible to distinguish between the contributions of
surface chemical spots or other processes invoked in the magnetospheric
accretion model to the observed profiles. As an example, we show in
Figure~\ref{fig:akelem} the behavior of the spectral lines
Fe\,{\footnotesize I}~5616, Ca\,{\footnotesize I}~6122,
Li\,{\footnotesize I}~6708, and the accretion diagnostic Na\,{\footnotesize I}
doublet at different orbital phases. As already mentioned above, at the
orbital phase $\phi$=0.946 the secondary component suffers from obscuration
and is only marginally detectable, probably due to the presence of clumpy dust
clouds in the circumbinary disk and/or inside the orbit of the system. The
fact that the secondary component becomes well visible at almost the same
orbital phase $\phi$= 0.950 in the observations obtained one year later
indicates that the dust obscuration in this system is varying from cycle to
cycle, with clouds probably existing at small scales as was already pointed out
by
\citet{Alencar2003}.
Linear polarimetric observations of
\citet{Manset2005}
reported the presence of significant variations of intrinsic percent
polarization and position angle indicating the presence of circumstellar
matter distributed in an asymmetric geometry. According to the authors, the
intrinsic polarization in AK\,Sco is one of the highest among the PMS binaries.
Variations were found to be periodical with a period equal to the
gravitational tidal period (half of the orbital period).

As already pointed out by
\citet{Alencar2003},
the line profiles of the Na\,{\footnotesize I} doublet are strongly perturbed
by the interaction with the disk. In Figure~\ref{fig:akew} we show the opposite
character of the variability of equivalent widths in both components. The
equivalent widths of photospheric lines in the primary component are
decreasing over the observed orbital phases range, while those for the
secondary component are increasing.

A number of Herbig\,Ae stars are known to exhibit $\delta$~Scuti-like
pulsations
\citep[e.g.,][]{Zwintz}.
Furthermore,
\citet{Gomezet}
reported the detection of a 1.3\,mHz ultra-low-frequency oscillation in the
ultraviolet light curve of AK\,Sco at periastron passage. The authors
suggested that this oscillation is most likely fed by the gravitational energy
released when the tails of the accretion stream spiraling onto each
star come into contact at periastron passage, enhancing the accretion
flow. Because HARPSpol observations  cover the orbital phases
corresponding to the periastron passage and are split into
subexposures taken at different angles of the quarter-wave retarder
plate, we checked for any changes in the line profile shape or radial
velocity (RV) shifts on the time scales of tens of minutes. We 
detect clear RV shifts with a decrease from 4.8\,km\,s$^{-1}$ at the 
phase 0.946 to 0.9\,km\,s$^{-1}$ at the phase 0.169. 
However, this RV variability corresponds to the RV changes expected for the 
orbital movement during the periastron passage. No typical feature 
characteristic for the pulsational variability is detected at any orbital phase.

Not much can be reported on the short-term spectral variability of HD\,95881
on a timescale of about 40--45 minutes, as no line profile or RV
variation were detected.


\section{Discussion}

Our detections of weak magnetic fields in two Herbig\,Ae stars confirm the
conclusion of
\citet{Hubrig2015}
that the previously accepted low incidence (5--10\%) of magnetic Herbig\,Ae
stars can be explained by the weakness of these fields and the limited
accuracy of the published measurements. The magnetic field was previously
searched but not detected in AK\,Sco by
\citet{Alecian2013}
using Echelle Spectro- PolArimetric Device for the Observation of Stars (ESPaDOnS)
observations. To our knowledge, the presence of a magnetic
field in HD\,95881 is reported in this work for the first time.
\citet{Wade}
used a single low-resolution ($R=1560$) spectropolarimetric observation of
this star obtained with FORS\,1 at the VLT to search for the presence of a
longitudinal magnetic field. Using different spectral regions, the most
accurate measurement, $\left<B_{\mathrm z}\right>=-20\pm42$\,G, was achieved
for the whole spectrum. We should keep in mind that the longitudinal
magnetic field is defined as the disk-integrated magnetic field component
along the line of sight and therefore shows a strong dependence on the
viewing angle of the observer, i.e., on the rotation angle of the star. The
limitations set by the strong geometric dependence of the longitudinal
magnetic field are usually overcome by repeating observations several times,
so as to sample various rotation phases, hence various aspects of the
magnetic field.

The double-lined system AK\,Sco appears to be of special interest for a
number of reasons: 
\citet{Czekala}
estimated the age of 18$\pm$1\,Myr, fully in line with its membership in the
Upper Centaurus--Lupus association, but surprisingly old for it to still host
a gas-rich disk. Recently, 
\citet{Donaldson}
measured the ages of PMS stars in the Upper Scorpius OB association and showed
that stars with disks have an older mean isochronal age than stars without
disks. The authors concluded that evolutionary effects in young stars can
affect their apparent ages. It is not clear whether the presence of a magnetic
field of AK\,Sco can influence the derived age.

The finding of the line intensity variability over the fraction of the orbital
period in AK\,Sco is intriguing and it is important to understand whether it
is caused by the star--disk interaction or by the inhomogeneous surface element
distribution.

Knowledge of the magnetic field structure combined with the determination of
the chemical composition are indispensable when constraining theories on star
formation and magnetospheric accretion in intermediate-mass stars. As of
today, magnetic phase curves (i.e.,\ the dependence of the magnetic field
strength on the rotation phase) have been obtained only for two Herbig\,Ae/Be
stars, V380\,Ori
\citep{Alecian2009}
and HD\,101412
\citep{HD101412}.
Furthermore, only very few close spectroscopic binaries with orbital periods
below 20\,days are known among Herbig\,Ae stars 
\citep{Duchene}.
A search for magnetic fields and the determination of their geometries in
close binary systems will play an important role for understanding of the
mechanisms that can be responsible for the magnetic field generation.

\acknowledgments

Based on observations made with ESO Telescopes at the La Silla Paranal
Observatory under programme IDs 097.C-0277(A) and 099.C-0081(A).
This work has made use of the VALD database, operated at Uppsala
University, the Institute of Astronomy RAS in Moscow, and the University of
Vienna.
M.A.P.\ acknowledges the Basic Research Program of the Presidium Of the Russian
Academy of Sciences P-28. 
N.A.D.\ acknowledges financial support by RFBR according to the research
project 18-02-00554.
We thank J.F.~Gonz\'alez for useful comments.

\vspace{5mm}
\facilities{ESO:3.6m}



\end{document}